\documentclass[preprint,aps,showpacs,preprintnumbers,amsmath,amssymb,nofootinbib,superscriptaddress]{revtex4}

\usepackage{graphicx}
\usepackage{dcolumn}
\usepackage{bm}
\usepackage{amsmath,amssymb}

\begin{document}

\preprint{YITP-08-2}
\preprint{KUNS-2148}
\preprint{KEK-CP-210}

\title{
  Nucleon sigma term and strange quark content from lattice QCD with exact
  chiral symmetry
}

\newcommand{\GUAS}{
  School of High Energy Accelerator Science,
  The Graduate University for Advanced Studies (Sokendai),
  Tsukuba 305-0801, Japan
}
\newcommand{\KEK}{
  High Energy Accelerator Research Organization (KEK),
  Tsukuba 305-0801, Japan
}

\author{
H. Ohki$^{1,2}$,
H. Fukaya$^3$,
S. Hashimoto$^{4,5}$,
T. Kaneko$^{4,5}$, \\
H. Matsufuru$^4$,
J. Noaki$^{4}$,
T. Onogi$^2$, 
E. Shintani$^4$,
N. Yamada$^{4,5}$
}

\affiliation{
$^1$  Department of Physics, 
      Kyoto University, Kyoto 606-8501, Japan,\\
$^2$  Yukawa Institute for Theoretical Physics, 
      Kyoto University, Kyoto 606-8502, Japan,\\
$^3$  The Niels Bohr Institute, The Niels Bohr International Academy 
      Blegdamsvej 17 DK-2100 Copenhagen, Denmark,\\
$^4$  High Energy Accelerator Research Organization (KEK), 
      Tsukuba 305-0801, Japan,\\
$^5$  School of High Energy Accelerator Science, 
      The Graduate University for Advanced Studies (Sokendai), 
      Tsukuba 305-0801, Japan,\\
}

\collaboration{JLQCD collaboration}

\begin{abstract}
  We calculate the nucleon sigma term in two-flavor lattice QCD utilizing the
  Feynman-Hellman theorem.
  Both sea and valence quarks are described by the overlap fermion
  formulation, which preserves exact chiral and flavor symmetries on the
  lattice.
  We analyse the lattice data for the nucleon mass using the analytical
  formulae derived from the baryon chiral perturbation theory.
  From the data at valence quark mass set different from sea quark mass, we
  may extract the sea quark contribution to the sigma term, which corresponds
  to the strange quark content.
  We find that the strange quark content is much smaller than the previous
  lattice calculations and phenomenological estimates.
\end{abstract}

\maketitle

\section{Introduction}
A piece of information on the nucleon structure can be extracted from
its quark mass dependence.
Nucleon sigma term $\sigma_{\pi N}$ characterizes the effect of finite
quark mass on the nucleon mass. 
Up to non-analytic and higher order terms, the nucleon mass is written 
as $M_N=M_0+\sigma_{\pi N}$, where $M_0$ is the nucleon mass in the chiral
limit.
The exact definition of $\sigma_{\pi N}$ is given by the form of a
scalar form factor of the nucleon at zero recoil as
\begin{equation}
  \label{eq:piNsigma}
  \sigma_{\pi N}= m_{ud} 
  \left( \langle N | \bar{u}u+\bar{d}d | N \rangle 
       -V\langle 0 | \bar{u}u+\bar{d}d | 0 \rangle 
  \right)
\end{equation}
where $m_{ud}$ denotes degenerate up and down quark mass.
The second term in the parenthesis represents a subtraction of the vacuum
contribution, and $V$ is the (three-dimensional) physical volume
\footnote{
  The nucleon state $|N(p)\rangle$ is normalized as
  $\langle N (p)|N (p^\prime)\rangle= (2\pi)^3 \delta^{(3)}(p-p^\prime)$.
  In (\ref{eq:piNsigma}) we omit the momentum argument for the nucleon,
  since we do not consider finite momentum insertion in this paper.
}.
For the sake of simplicity we represent the vacuum subtracted matrix
element $\langle N|\bar{q}q|N\rangle-V\langle 0|\bar{q}q|0\rangle$
by $\langle N|\bar{q}q|N\rangle$ in what follows.
($q$ represents a quark field: up ($u$), down ($d$), or strange ($s$).)
Note that the sigma term is renormalization group invariant, since the
renormalization factor cancels between the quark mass $m_q$ and the
scalar operator $\bar{q}q$. 

While the up and down quarks contribute to $\sigma_{\pi N}$ both as
valence and sea quarks, the strange quark appears only as a sea quark
contribution. 
As a measure of the strange quark content of the nucleon, the $y$ parameter
\begin{equation}
  \label{eq:y}
  y \equiv 
  \frac{2\langle N | \bar{s}s  | N \rangle}
  {\langle N | \bar{u}u +\bar{d}d | N \rangle},
\end{equation}
is commonly introduced.
Besides characterizing the purely sea quark content of the nucleon,
which implies a clear distinction from the quark model picture of hadrons,
this parameter plays an important role to determine the detection rate
of possible neutralino dark matter in the supersymmetric extension of
the Standard Model  
\cite{Griest:1988yr,Griest:1988ma,Bottino:1999ei,
  Ellis:2003cw,Ellis:2005mb,Baltz:2006fm,Ellis:2008hf}.
Already with the present direct dark matter search experiments one may
probe a part of the MSSM model parameter space, and new experiments
such as XMASS and SuperCDM will be able to improve the sensitivity by
2--3 orders of magnitude. 
Therefore, a precise calculation of the $y$ parameter 
(or equivalently another parameter
$f_{T_s}\equiv m_s\langle N|\bar{s}s|N \rangle/M_N$)
will be important for excluding or proving the neutralino dark matter
scenario. 

Phenomenologically, the sigma term can be related to the $\pi N$
scattering amplitude at a certain kinematical point, {\it i.e.} the
so-called Cheng-Dashen point $t=+2m_\pi^2$ \cite{Cheng:1970mx}.
Its value is in the range $\Sigma_{CD}=70\sim 90$~MeV \cite{Pavan:2001wz}.
After the corrections for the finite value of $t$, which amounts to
$-15$~MeV \cite{Gasser:1990ce}, one 
obtains $\sigma_{\pi N}=55\sim 75$~MeV.
On the other hand, the octet breaking of the nucleon mass, or the matrix
element $\langle N|\bar{u}u+\bar{d}d-2\bar{s}s|N\rangle$, can be evaluated
from the baryon mass spectrum.
At the leading order of Chiral Perturbation Theory (ChPT), the
value of the corresponding sigma term is $\hat{\sigma}\simeq 26$~MeV,
while the heavy baryon ChPT (BChPT) gives 
$\hat{\sigma}=36\pm 7$~MeV~\cite{Borasoy:1996bx}.
The difference between $\sigma_{\pi N}$ and $\hat{\sigma}$ is understood as
the strange quark contribution; algebraically the relation is 
$\sigma_{\pi N} = \hat{\sigma}/(1-y)$.
Then, one obtains a large value of $y$: y=0.3--0.6.
(The value of $y$ is even larger than the estimate $y\simeq 0.2$ in
\cite{Gasser:1990ce}, because of the more recent experimental data
\cite{Pavan:2001wz}.) 
For other phenomenological estimates, see, {\it e.g.} \cite{Bernard:2007zu}.
Such large values of $y$ cannot be understood within the valence quark
picture, hence raises a serious problem in the understanding of the nucleon
structure. 
We note however that the analysis within chiral effective theories suffers
from significant uncertainties of the low energy constants, especially at
higher orders.

Using lattice QCD, one can in principle calculate the nucleon sigma term
without involving any model parameters,
since lattice calculation for a wide range of quark masses in the
chiral regime offers essential information on the low energy constants
which cannot be determined by experimental data alone.
Furthermore, it is possible to determine the valence and sea quark
contributions separately.
A direct method to extract them is to calculate three-point functions
on the lattice including an insertion of the scalar operator.
It can also be done in an indirect way by analyzing the quark mass dependence
of the nucleon mass for valence and sea quarks separately.
Obviously, the dynamical fermion simulations are necessary to extract the
disconnected contributions in the indirect method.

Previous lattice results were
$\sigma_{\pi N}$ = 40--60~MeV, $y$ = 0.66(15) \cite{Fukugita:1995ba},
and
$\sigma_{\pi N}$ = 50(3)~MeV, $y$ = 0.36(3) \cite{Dong:1996ec} 
within the quenched approximation, 
while a two-flavor QCD calculation \cite{Gusken:1998wy} gave
$\sigma_{\pi N}$ = 18(5)~MeV and $y$ = 0.59(13).
There are apparent puzzles in these results: 
firstly the strange quark content due to the disconnected diagram (the value
of $y$) is unnaturally large compared to the up and down contributions that
contain the connected diagrams too.
Secondly the values of the sigma term in the quenched and unquenched
calculations are rather different, which might also imply significant effects
of quark loops in the sea.

Concerning the first point, it was pointed out that using the Wilson-type
fermions, which violate the chiral symmetry on the lattice, the sea quark
mass dependence of the additive mass renormalization and lattice spacing can
give rise to a significant uncertainty in the sea quark content
\cite{Michael:2001bv}.
Unfortunately, after subtracting this contamination the unquenched result
has large statistical error, $y=-0.28(33)$.
In the present work, we remove this problem by explicitly maintaining
exact chiral symmetry on the lattice for both sea and valence sectors,
as described below. 

The second puzzle may be resolved by incorporating an enhancement due
to pion loops.
Within BChPT at $O(p^3)$ or $O(p^4)$, a curvature is expected in the
quark mass dependence of the nucleon, hence the sigma term, a derivative
of $M_N$ in terms of $m_q$, increases towards the chiral limit.

An analysis using existing lattice data of two-flavor QCD
with $m_\pi>550$~MeV
by CP-PACS \cite{Ali Khan:2001tx},
JLQCD \cite{Aoki:2002uc}, and QCDSF \cite{Ali Khan:2003cu}
yields $\sigma_{\pi N}=48\pm 5 ^{+\ 9}_{-12}$~MeV
\cite{Procura:2003ig,Procura:2006bj},
which is slightly smaller than but is
still consistent with the phenomenological analysis.
A more recent lattice data by the ETM Collaboration
with $m_\pi$ = 300--500~MeV
in two-flavor QCD reported a higher value $\sigma_{\pi N}$ = 67(8)~MeV
\cite{Alexandrou:2008tn}.
Such an analysis for the disconnected contribution to extract the
strange content is yet to be done, which is another main point of
this work.

In this work, we analyze the data of the nucleon mass obtained from a
two-flavor QCD simulation employing the overlap fermion \cite{Aoki:2008tq}.
(For other physics results from this simulation, we refer
\cite{Matsufuru:2007uc} and references therein.)
The overlap fermion \cite{Neuberger:1997fp,Neuberger:1998wv} preserves
exact chiral symmetry on the lattice, and there is no problem of the
additive mass shift of the scalar density operator, that was a main
source of the large systematic error in the previous calculations of the
sigma term.
We use the overlap fermion to describe both the sea and valence quarks.
Statistically independent ensembles of gauge configurations are
generated at six different sea quark masses; 
the nucleon mass is measured for various valence quark masses on each
of those gauge ensembles.
Therefore, we are able to analyze the valence and sea quark mass
dependence independently to extract the connected and disconnected
contributions. 
An estimate of the strange quark content can thus be obtained in 
two-flavor QCD.
In the analysis, we use the partially quenched BChPT, which
corresponds to the lattice calculations with valence quark masses
taken differently from the sea quark masses. 
Therefore, the enhancement of the sigma term towards the chiral limit
is incorporated.
Since the two-flavor QCD calculation cannot avoid the
systematic error due to the neglected strange sea quarks,
our result should not be taken as a final result from lattice QCD.
Nevertheless our study with exact chiral symmetry reveals
the underlying systematic effects in the calculation of the nucleon
sigma term, especially in the extraction of its disconnected
contribution.
It therefore provides a realistic test case,
which will be followed by the 2+1-flavor calculations in the near
future
\footnote{
  For a very recent result from 2+1-flavor QCD, see
  \cite{WalkerLoud:2008bp}.
}.

Our paper is organized as follows.
In Section \ref{sec:Method}, we introduce the basic methods to calculate
the nucleon sigma term.
Our simulation set-up is described in Section~\ref{sec:Simulation}.
Then, in Section~\ref{sec:BChPT}, we describe the BChPT fit to obtain the
sigma term. 
In Section~\ref{sec:PQChPT}, we study the sea quark content of the nucleon
from PQChPT.
In Section~\ref{sec:Discussion}, we compare our results with previous
calculations and discuss the origin of the discrepancy.
Our conclusion is given in Section~\ref{sec:Summary}.

\section{Method for calculating Nucleon sigma term}
\label{sec:Method}
The matrix element defining the nucleon sigma term (\ref{eq:piNsigma})
can be related to the quark mass dependence of the nucleon mass using
the Feynman-Hellman theorem. 
Consider a two-point function of the nucleon interpolating operator
$O_N(t,\vec{x})$ 
\begin{equation}
  G(t) \equiv \int d^3\vec{x}  
  \langle 0 | O_N(t,\vec{x}) O^{\dagger}_N(0,\vec{0})| 0 \rangle
  = Z^{-1} \int {\cal D}A_{\mu} \prod_q ({\cal D}q {\cal D}\bar{q})
  \,
  O_N(t,\vec{x}) O^{\dagger}_N(0,\vec{0})
  \, e^{-S}
\end{equation}
with the QCD action $S$ defined by the gluon field strength $F_{\mu\nu}$ 
and the quark field $q$ as
\begin{equation}
  \label{eq:action}
  S= \int d^4 x    
  \left\{
    \frac{1}{2g^2}\mathrm{Tr} F_{\mu\nu}^2  + \sum_{q=u,d} \bar{q} (D+m_q) q
  \right\}
\end{equation}
and the partition function $Z$.
The sum in (\ref{eq:action}) runs over flavors ($q$ = $u$ and $d$)
according to the underlying two-flavor theory ($N_f$ = 2). 
By taking a partial derivative with respect to a valence quark mass
$m_{\mathrm{val}}$ or a sea quark mass $m_{\mathrm{sea}}$
corresponding to the degenerate $u$ and $d$ quark masses 
$m_{ud}$ ($=m_u=m_d$), we obtain
\begin{eqnarray}
  \label{eq:deriv_val}
  \frac{\partial G(t)}{\partial m_{\mathrm{val}}}
  &=& 
  - \int\! d^3\vec{x}
  \left\langle 0 \left|  
      O_N(t,\vec{x})O^{\dagger}_N(0,\vec{0})
      \left[\int\! d^4y \sum_{q=u,d}(\bar{q}q)(y) \right]
  \right| 0\right\rangle_{\mathrm{conn}},
  \\ 
    \label{eq:deriv_sea}
  \frac{\partial G(t)}{\partial m_{\mathrm{sea}}}
  &=& 
  - \int\! d^3\vec{x}
  \left\langle 0 \left|
      O_N(t,\vec{x})O^{\dagger}_N(0,\vec{0})
      \left[\int\! d^4y \sum_{q=u,d} (\bar{q}q)(y)\right]
      \right| 0 \right\rangle_{\mathrm{disc}}
    \nonumber\\
  & & 
  + G(t) 
  \left\langle 0 \left|
      \int\! d^4y \sum_{q=u,d}(\bar{q}q)(y)
  \right| 0 \right\rangle.
\end{eqnarray} 
The subscripts ``conn'' and ``disc'' on the expectation values indicate that
only the connected or disconnected quark line contractions are evaluated,
respectively. 

Dividing the integration region of $t_y$, a temporal component of $y$, into
three parts, {\it i.e.} $t_y<0$, $0<t_y<t$, and $t<t_y$, and inserting the
complete set of states between the operators, one can express 
$G(t)$, 
$\partial G(t)/\partial m_{\mathrm{val}}$, and
$\partial G(t)/\partial m_{\mathrm{sea}}$ in terms of
matrix elements. 
Comparing the leading contribution at large $t$ behaving as 
$t\exp\left(-M_N t\right)$ with $M_N$ the nucleon mass,  
we obtain the relations
\begin{eqnarray}
  \label{eq:deriv_mN_val}
  \frac{\partial M_N}{\partial m_{\rm val}} 
  &=& \langle N | (\bar{u}u+\bar{d}d) | N\rangle_{\rm conn},
  \\
  \label{eq:deriv_mN_sea}
  \frac{\partial M_N}{\partial m_{\rm sea}} \
  &=& \langle N | (\bar{u}u+\bar{d}d) | N\rangle_{\rm disc},
\end{eqnarray}
Note that the short-hand notation
to omit the term $-V\langle 0|(\bar{q}q)|0\rangle$ applies
only for the disconnected piece.

This derivation of the Feynman-Hellman theorem does not assume anything about
the renormalization scheme nor the regularization scheme. 
The contact terms in (\ref{eq:deriv_val}) and (\ref{eq:deriv_sea}) are
irrelevant for the formulas (\ref{eq:deriv_mN_val}) and
(\ref{eq:deriv_mN_sea}), since only the long-distance behavior of the
correlators is used.

In the present study we exploit this indirect method to extract the matrix
elements corresponding to the nucleon sigma term.

Another possible method to calculate the nucleon sigma term is to directly
calculate the matrix element from three-point functions with an
insertion of the scalar density operator $(\bar{u}u+\bar{d}d)(x)$, as
carried out, {\it e.g.} in \cite{Fukugita:1995ba,Dong:1996ec} in the
quenched approximation.
In principle, it gives a mathematically equivalent quantity to the
indirect method, provided that the indirect method is applied with data at
sufficiently many sets of $(m_{val},m_{sea})$ so that the derivatives
are reliably extracted.
The order of the derivative and the path integral does not make any
difference at finite lattice spacing and volume.
Numerical difference could arise only from the statistical error and
the systematic error in the fit of the data.
A practical advantage of the indirect method is that the sum over
the position of $(\bar{u}u+\bar{d}d)(x)$ is automatic,  
whereas in the direct method it must be taken explicitly to improve
statistical accuracy.
On the other hand, the direct method is more flexible, as one can
take arbitrary quark masses for the ``probe quark'' to make a
disconnected loop from the $(\bar{u}u+\bar{d}d)(x)$ operator, while in
the indirect method the probe quark mass is tied to the sea
quark mass.
Therefore, we can only estimate the strange quark content from the
calculation done at the strange quark mass equal to the sea quark mass
as
$\partial M_N/\partial m_{\mathrm{sea}}|_{m_{\mathrm{val}}=m_{\mathrm{sea}}=m_s} =
2\langle N|\bar{s}s|N\rangle_{\mathrm{disc}}$.

\section{Lattice simulation}
\label{sec:Simulation}

We make an analysis of the nucleon mass using the lattice
data obtained on two-flavor QCD configurations generated with
dynamical overlap fermions \cite{Aoki:2008tq}.
The lattice size is $16^3\times 32$, which roughly corresponds to the
physical volume (1.9~fm)$^3\times$(3.8~fm) with the lattice spacing 
determined through the Sommer scale $r_0$ as described below.
The overlap fermion is defined with the overlap-Dirac operator
\cite{Neuberger:1997fp,Neuberger:1998wv} 
\begin{equation}
  \label{eq:overlap}
  D(m_q)=\left(m_0+\frac{m_q}{2}\right)
  + \left(m_0-\frac{m_q}{2}\right)\gamma_5 \mathrm{sgn}\left[H_W(-m_0)\right]
\end{equation}
for a finite (bare) quark mass $m_q$.
The kernel operator $H_W(-m_0)\equiv\gamma_5D_W(-m_0)$ is constructed
from the conventional Wilson-Dirac operator $D_W(-m_0)$ at
a large negative mass $-m_0$. (We set $m_0=1.6$ in this work.)
For the gluon part, the Iwasaki action is used at $\beta$ = 2.30 together with 
unphysical Wilson fermions and associated twisted-mass ghosts
\cite{Fukaya:2006vs} introduced to suppress unphysical near-zero modes
of $H_W(-m_0)$. 
With these extra terms, the numerical operation for applying the
overlap-Dirac operator (\ref{eq:overlap}) is substantially reduced.
Furthermore, since the exact zero eigenvalue is forbidden, the global
topological charge $Q$ is preserved during the molecular dynamics
evolution of the gauge field.  
Our main runs are performed at the trivial topological sector $Q=0$.
For each sea quark mass listed below, we accumulate 10,000
trajectories; the calculation of the nucleon mass is done at every
20 trajectories, thus we have 500 samples for each $m_{\mathrm{sea}}$.  
For more details of the configuration generation, we refer to
\cite{Aoki:2008tq}. 

\begin{table}[tbp]
  \begin{center}
    \begin{tabular}{cccc}
      \hline
      $ am_{\mathrm{sea}}$ & $a$ [fm] & $m_{\pi}$ [GeV] & $m_{\pi} L$ \\
      \hline
      0.015 & 0.1194(15) & 0.2880(18) & 2.8\\
      0.025 & 0.1206(18) & 0.3671(13) & 3.5\\
      0.035 & 0.1215(15) & 0.4358(13) & 4.2\\
      0.050 & 0.1236(14) & 0.5217(13) & 5.0\\
      0.070 & 0.1251(13) & 0.6214(11) & 6.0\\
      0.100 & 0.1272(12) & 0.7516(14) & 7.2\\
      \hline
    \end{tabular}
    \caption{Lattice spacing and pion mass calculated for each sea
      quark mass.}
    \label{tab:param}
  \end{center}
\end{table}

For the sea quark mass $am_{\mathrm{sea}}$ we take six values: 0.015,
0.025, 0.035, 0.050, 0.070, and 0.100 that cover the mass range
$m_s/6$--$m_s$ with $m_s$ the physical strange quark mass.
Analysis of the pion mass and decay constant on this data set is found
in \cite{Noaki:2008iy}.

The lattice spacing determined through the Sommer scale $r_0$
of the static quark potential slightly depends on the sea quark mass;
the numerical results are listed in Table~\ref{tab:param}.
Extrapolating to the chiral limit, we obtain $a$ = 0.118(2)~fm
assuming the physical value $r_0=0.49$~fm.
In the following analysis, we use this value to convert the lattice
results to the physical unit.

The two-point functions, from which the nucleon mass is extracted, are
constructed from quark propagators described by the overlap fermion.
In order to improve the statistical accuracy, we use the low-mode
preconditioning technique \cite{DeGrand:2004qw}, {\it i.e.} the piece
of the two-point function made of the low modes of the overlap-Dirac
operator is averaged over many source points.
In our case, the source points are set at the origin on each time
slice and averaged over different time slices with 50 chiral pairs 
of low modes.
For the source to solve the quark propagator, we take a smeared source
defined by a function 
$\phi(|\vec{x}|)\propto\exp(-A|\vec{x}|)$ with a fixed $A$ = 0.40.
We then calculate the smeared-local two-point correlator
and fit the data with a single exponential function after averaging
over forward and backward propagating states in time.
The statistical error is estimated using the standard jackknife method with
a bin size of 10 samples, which corresponds to 200 trajectories.
In the calculation of the nucleon mass, we take the valence quark
masses $am_{\mathrm{val}}$ = 
0.015, 0.025, 0.035, 0.050, 0.060, 0.070, 0.080, 0.090, and 0.100.

\begin{figure}[tbp]
  \begin{center}
      \includegraphics[width=10cm]{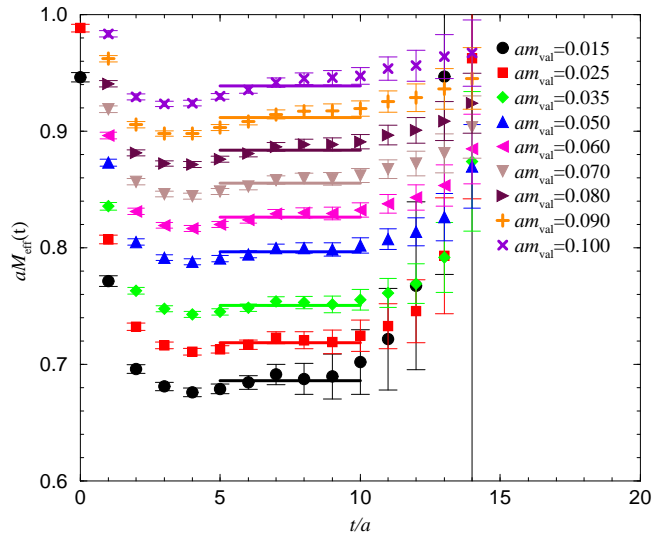}
  \end{center}
  \caption{Effective mass of the smeared-local nucleon correlator.
    Data are shown for various degenerate valence quark masses
    $m_{\mathrm{val}}$ at a fixed sea quark mass $am_{\mathrm{sea}}=0.035$.}
   \label{fig:meff}
\end{figure}%

Figure~\ref{fig:meff} shows an effective mass plot of the nucleon 
at $am_{\mathrm{sea}}=0.035$.
Data are shown for the nucleon made of degenerate valence
quarks at the nine available masses.
We find a good plateau for all sea and valence quark mass
combinations; the fit with a single exponential function is made in
the range [5,10].
The fitted results are shown by thick horizontal lines in
Figure~\ref{fig:meff} and summarized in Table~\ref{tab:mass}. 

\begin{table}[tbp]
\begin{tabular}{cccc}
\hline
  $am_{\mathrm{sea}}$ &  $am_{\mathrm{val}}$ & $am_{PS}$ & $aM_N$ \\ 
\hline
 0.015  &   0.015 &\ \ 0.1729(12)  &   0.6647(59)\\
        &   0.025 &\ \ 0.2210(10)  &   0.7038(47)\\
        &   0.035 &\ \ 0.25998(95) &   0.7381(44)\\
        &   0.050 &\ \ 0.30966(93) &   0.7858(43)\\
        &   0.060 &\ \ 0.33918(96) &   0.8164(43)\\
        &   0.070 &\ \ 0.3668(10)  &   0.8460(44)\\
        &   0.080 &\ \ 0.3929(11)  &   0.8750(45)\\
        &   0.090 &\ \ 0.4180(12)  &   0.9033(47)\\
        &   0.100 &\ \ 0.4423(15)  &   0.9312(48)\\
\hline
 0.025  &   0.015 &\ 0.17185(99)  &   0.6597(60)\\
        &   0.025 &\ 0.21990(81)  &   0.6960(45)\\
        &   0.035 &\ 0.25906(76)  &   0.7300(40)\\
        &   0.050 &\ 0.30900(72)  &   0.7780(37)\\
        &   0.060 &\ 0.33864(72)  &   0.8084(37)\\
        &   0.070 &\ 0.36625(75)  &   0.8380(39)\\
        &   0.080 &\ 0.39230(81)  &   0.8667(38)\\
        &   0.090 &\ 0.41711(92)  &   0.8946(39)\\
        &   0.100 &\ 0.4409(11)   &   0.9220(42)\\
\hline
 0.035  &   0.015 &\ 0.17299(96)  &   0.6859(73)\\
        &   0.025 &\ 0.22168(84)  &   0.7186(45)\\
        &   0.035 &\ 0.26111(81)  &   0.7505(35)\\
        &   0.050 &\ 0.31136(82)  &   0.7966(30)\\
        &   0.060 &\ 0.34124(85)  &   0.8263(30)\\
        &   0.070 &\ 0.36917(90)  &   0.8553(30)\\
        &   0.080 &\ 0.39567(99)  &   0.8837(30)\\
        &   0.090 &\ 0.4211(11)   &   0.9115(32)\\
        &   0.100 &\ 0.4456(14)   &   0.9388(33)\\
\hline
\end{tabular}
\hspace*{4mm}
\begin{tabular}{cccc}
\hline
  $am_{\mathrm{sea}}$ &  $am_{\mathrm{val}}$ & $am_{PS}$ & $aM_N$ \\ 
\hline
 0.050  &   0.015 &\ 0.17402(87)  &   0.6895(58)\\
        &   0.025 &\ 0.22279(82)  &   0.7274(43)\\
        &   0.035 &\ 0.26218(80)  &   0.7606(39)\\
        &   0.050 &\ 0.31228(80)  &   0.8072(38)\\
        &   0.060 &\ 0.34199(82)  &   0.8369(40)\\
        &   0.070 &\ 0.36967(84)  &   0.8657(39)\\
        &   0.080 &\ 0.39583(93)  &   0.8937(41)\\
        &   0.090 &\ 0.4208(10)   &   0.9211(43)\\
        &   0.100 &\ 0.4449(13)   &   0.9480(45)\\
\hline
 0.070  &   0.015 &\ 0.17512(76)  &   0.6870(63)\\
        &   0.025 &\ 0.22444(65)  &   0.7259(44)\\ 
        &   0.035 &\ 0.26399(62)  &   0.7610(39)\\
        &   0.050 &\ 0.31414(64)  &   0.8098(37)\\ 
        &   0.060 &\ 0.34388(68)  &   0.8407(37)\\ 
        &   0.070 &\ 0.37166(75)  &   0.8705(38)\\ 
        &   0.080 &\ 0.39800(88)  &   0.8996(39)\\
        &   0.090 &\ 0.4232(11)   &   0.9280(41)\\ 
        &   0.100 &\ 0.4477(14)   &   0.9559(42)\\
\hline
 0.100  &   0.015 &\ 0.17663(71)  &   0.7040(65)\\
        &   0.025 &\ 0.22563(61)  &   0.7419(44)\\
        &   0.035 &\ 0.26548(58)  &   0.7761(37)\\
        &   0.050 &\ 0.31619(56)  &   0.8236(35)\\
        &   0.060 &\ 0.34621(57)  &   0.8536(34)\\
        &   0.070 &\ 0.37418(61)  &   0.8827(34)\\
        &   0.080 &\ 0.40061(70)  &   0.9110(35)\\
        &   0.090 &\ 0.42587(87)  &   0.9387(36)\\
        &   0.100 &\ 0.4502(11)   &   0.9659(37)\\
\hline
\end{tabular}
\caption{Numerical results for the pseudo-scalar meson mass 
and  the nucleon mass for each sea and valence quark masses.}
\label{tab:mass}
\end{table}

\section{Analysis of the unitary points with baryon chiral
  perturbation theory}
\label{sec:BChPT} 

\subsection{Naive fits with BChPT}

In this section, we analyze the lattice data taken at the unitary
points, {\it i.e.} sea and valence quarks are degenerate.
In this case, the conventional baryon chiral perturbation theory
(BChPT) \cite{Jenkins:1990jv} for two flavors is a valid framework to
describe the quark mass dependence of the nucleon.
It develops a non-analytic quark mass dependence and leads to the
enhancement of the nucleon sigma term near the chiral limit.

In BChPT, the nucleon mass is expanded in terms of the light quark mass
or equivalently pion mass squared $m_\pi^2$.
We follow the analysis done in \cite{Procura:2003ig}.
The expression for the nucleon mass $M_N$ to the order
$\mathcal{O}(p^3)$ has a form 
\begin{eqnarray}
  \label{eq:p^3}
  M_N &=& M_0 -4c_1 m_\pi^2
        -\frac{3g_A^2}{32\pi f_\pi^2}m_\pi^3  
     +\left[ 
	  e_1^r(\mu)-\frac{3g_A^2}{64\pi^2f_\pi^2 M_0}
	  \left( 1+2\log{\frac{m_\pi}{\mu}} \right)
	 \right] m_\pi^4 \nonumber\\
    &  & +\frac{3g_A^2}{256\pi f_\pi^2 M_0^2} m_\pi^5,
\end{eqnarray}	
and that to $\mathcal{O}(p^4)$ is
\begin{eqnarray}
  \label{eq:p^4}
  M_N &=& M_0 -4c_1 m_\pi^2
        -\frac{3g_A^2}{32\pi f_\pi^2}m_\pi^3  \nonumber \\
    & & +\left[ 
	  e_1^r(\mu)-\frac{3}{64\pi^2f_\pi^2}
	  \left( \frac{g_A^2}{M_0}-\frac{c_2}{2}\right)
	  -\frac{3}{32\pi^2f_\pi^2}
	  \left(\frac{g_A^2}{M_0}-8c_1+c_2+4c_3 \right)
	  \log{\frac{m_\pi}{\mu}} 
	 \right] m_\pi^4 \nonumber \\
    & & +\frac{3g_A^2}{256\pi f_\pi^2 M_0^2} m_\pi^5.
\end{eqnarray}	
There are many parameters involved in these expressions.
First of all, $M_0$ is the nucleon mass in the chiral limit and
$f_\pi$ is the pion decay constant.
The constant $g_A$ describes the nucleon axial-vector coupling.
Its experimental value determined by the neutron $\beta$ decay
is $g_A=1.270(3)$ \cite{Yao:2006px}.
The parameters $c_1$, $c_2$, and $c_3$ are low energy constants (LECs)
at $\mathcal{O}(p^2)$; their phenomenological values are
$c_1=-0.9^{+0.2}_{-0.5}$~GeV$^{-1}$,
$c_2=3.3\pm 0.2$~GeV$^{-1}$, and
$c_3=-4.7^{+1.2}_{-1.0}$~GeV$^{-1}$
(for a summary, see, for example \cite{Bernard:2007zu}).
In the fit using (\ref{eq:p^4}) discussed below, we fix 
$(c_2,c_3)$ at two representative combinations,
(3.2~GeV$^{-1}$,$-3.4$~GeV$^{-1}$) and 
(3.2~GeV$^{-1}$,$-4.7$~GeV$^{-1}$), following the previous analysis 
\cite{Procura:2003ig}.
As given above, $c_2$ is rather well determined phenomenologically.
As for $c_3$, the value $-$3.4~GeV$^{-1}$ is consistent with empirical
nucleon-nucleon phase shifts, and the value $-$4.7~GeV$^{-1}$ is
the central value obtained from pion-nucleon scattering.
There is another parameter $e_1^r(\mu)$, which is a combination of the
$\mathcal{O}(p^4)$ LECs and is not well known phenomenologically.
Since $e_1^r(\mu)$ is scale dependent, we quote its value at $\mu$ =
1~GeV in the following.

In these formulae, the leading non-analytic quark mass dependence is
given by the term of $m_\pi^3$, while others ($m_\pi^4\log(m_\pi/\mu)$
and $m_\pi^5$) are suppressed by additional powers of $m_\pi/M_0$.
Therefore, we also consider a simplified fit function
\begin{equation}
  \label{eq:p^3+higher}
  M_N = M_0 -4c_1 m_\pi^2 -\frac{3g_A^2}{32\pi f_\pi^2}m_\pi^3 
  +  e_1^r(\mu)	 m_\pi^4 .
\end{equation}
When we analyse the partially quenched data set in the next section,
we utilize a formula that is an extension of this simplified fit
form. 
Therefore, a comparison of the simplified and the full fit functions
(\ref{eq:p^3}), (\ref{eq:p^4}) on the unitary data points provides a
good test of our analysis.

We carry out the BChPT fits of the lattice data using these functions.
The simplest fits are those with (\ref{eq:p^3+higher}).
Since the axial-coupling $g_A$ is very well known experimentally, we
attempt two options: (Fit 0a) a fit with fixed $g_A$ (=1.267), and
(Fit 0b) a fit with $g_A$ being dealt as a free parameter.
The fits using the $\mathcal{O}(p^3)$ formula (\ref{eq:p^3}) are called Fit I.
Again in this case, we attempt the fits with (Fit Ia) and without (Fit
Ib) fixing $g_A$.
For the fit using the $\mathcal{O}(p^4)$ formula (\ref{eq:p^3}), the
lattice data do not have enough sensitivity to determine many
parameters in the formula unless we fix $g_A$, $c_2$, and $c_3$.
As described above we choose $g_A$ = 1.267, $c_2$ = 3.2~GeV$^{-2}$,
and $c_3=-$3.4~GeV$^{-1}$ (Fit II) or $c_3=-$4.7~GeV$^{-1}$ (Fit III).
The pion decay constant is fixed at its physical value 92.4~MeV.

We use the lattice data at five quark masses 
$m_q$ = 0.025, 0.035, 0.050, 0.070, 0.100. 
The data point at the smallest quark mass $m_q$ = 0.015 is not
included in the fit in order to avoid large finite volume effect. 
A detailed discussion on the finite volume effect is given below.

\begin{figure}[tbp]
  \begin{center}
    \rotatebox{0}{
      \includegraphics[width=10cm,clip]{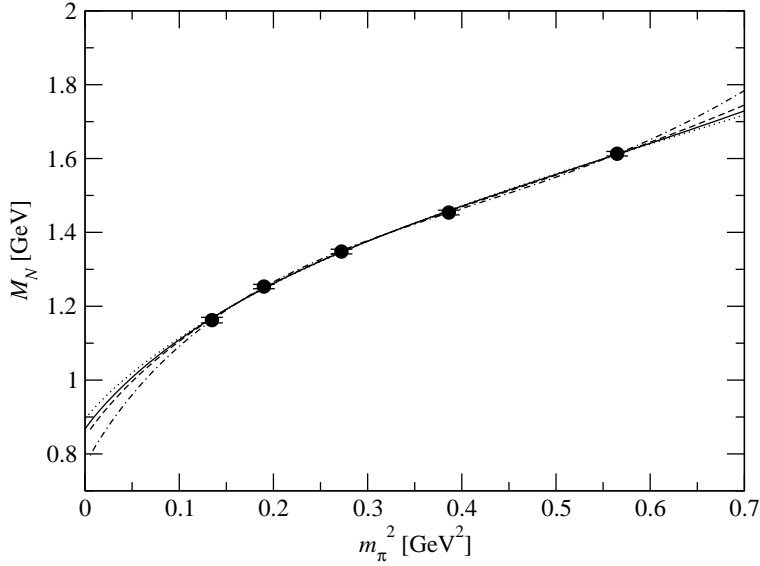}
    }
    \caption{
      BChPT fit of the nucleon mass for unitary points. 
      The solid, dot, dashed, dot-dashed curves represent the Fit 0a,
      Ia, II, and III, respectively.
    }
    \label{fig:ChPT}
  \end{center}
\end{figure}

\begin{table}[tbp]
\centering
\begin{tabular}{cccccccc}
\hline
   &   $M_0$[GeV] &  $c_1$[GeV$^{-1}$] & $e_1^r(\mu)$ 
   & $g_A$ &  $c_2$[GeV$^{-1}$] & $c_3$[GeV$^{-1}$] 
   & $\chi^2/{\mathrm{d.o.f.}}$ \\ 
\hline
Fit 0a &0.868(15) & $-$0.97(3) & 2.89(15)  &[1.267]& - & - 
&  0.89 \\
Fit 0b &0.753(106)& $-$1.59(56)& 6.7(3.6)  &1.81(42)  & - & - 
&  1.26 \\
Fit Ia &0.895(15) & $-$0.86(3) & 3.34(16)  &[1.267]& - & -  
&  1.39 \\
Fit Ib &0.748(104)& $-$1.72(59)&10.5(4.7)  &2.13(48)  & - & -  
&  0.88 \\
Fit II &0.846(13) & $-$1.04(2) & 2.05(11)  &[1.267] &[3.2] &[$-$3.4]
&  0.44 \\
Fit III &0.770(13) & $-$1.31(2) & 1.33(12) &[1.267] &[3.2] &[$-$4.7]
&  0.11 \\
\hline
\end{tabular}
\caption{ChPT fit of the nucleon mass using five unitary points 
  $m_q=$ 0.025, 0.035, 0.050, 0.070, and 0.100. 
  The values sandwiched as $[\cdots]$ mean the input in the fit.
}
\label{tab:ChPT}
\end{table}

Figure~\ref{fig:ChPT} shows the ChPT fits; the resulting fit
parameters are listed in Table~\ref{tab:ChPT}. 
The fit curves are drawn for Fit 0a, Ia, II and III in Figure~\ref{fig:ChPT}.
The lattice data show a significant curvature towards the chiral limit.
All of the fit functions with $g_A$ fixed to the experimental value
describe the data quite well.  
The results with $g_A$ a free parameter (Fit 0b and Ib) give an
important consistency check of BChPT, since the $m_\pi^3$ term is an
unique consequence of the pion loop effect in this framework.
The coupling $g_A$ is in fact non-zero and roughly consistent with the
experimental value within a large statistical error.
The nucleon mass in the chiral limit $M_0$ shows a significant
variation, especially when the Fit III is used.

\subsection{Finite volume corrections}
Since the spatial extent $L$ of the lattice is not large enough
($\sim$ 1.9~fm) for obtaining the baryon masses very accurately,
we need to estimate the systematic error due to the finite volume
effect. 

The finite volume correction can be calculated within BChPT, provided
that the quark mass is small enough to apply ChPT.
The nucleon mass $M_N(L)$ in a finite box of size $L^3$ is written as 
\cite{Ali Khan:2003cu} 
\begin{equation}
  \label{eq:FSE}
  M_N(L)-M_N(\infty)=\Delta_a + \Delta_b +\mathcal{O}(p^5),
\end{equation}
where $\Delta_a$ and $\Delta_b$ represent finite volume correction 
at order $p^3$ and $p^4$ respectively,
\begin{eqnarray}
  \label{eq:order3} 
  \Delta_a &=&  
  \frac{3g_A^2M_0 m_\pi^2}{16\pi^2 f_\pi^2}
  \int_0^\infty dx \sum_{\vec{n}} 
  K_0(L|n|\sqrt{M_0^2 x^2 +m_\pi^2(1-x)}),
  \\
  \label{eq:order4}
  \Delta_b &=& \frac{3m_\pi^4}{4\pi^2 f_\pi^2}
  \sum_{\vec{n}}\left[
    (2c_1-c_3)\frac{K_1(L|n|m_\pi)}{L|n|m_\pi}
    +c_2 \frac{K_2(L|n|m_\pi)}{(L|n|m_\pi)^2}
  \right].
\end{eqnarray}
Here, the functions $K_0(x)$, $K_1(x)$ and $K_2(x)$ are the modified
Bessel functions, which asymptotically behave as $\exp(-x)$ for large
$x$. 
The sum runs over a three dimensional vector $\vec{n}$ of integer
components, and $|n|$ denotes $\sqrt{\vec{n}^2}$.

\begin{figure}[tbp]
  \centering
  \rotatebox{0}{
    \includegraphics[width=10cm,clip]{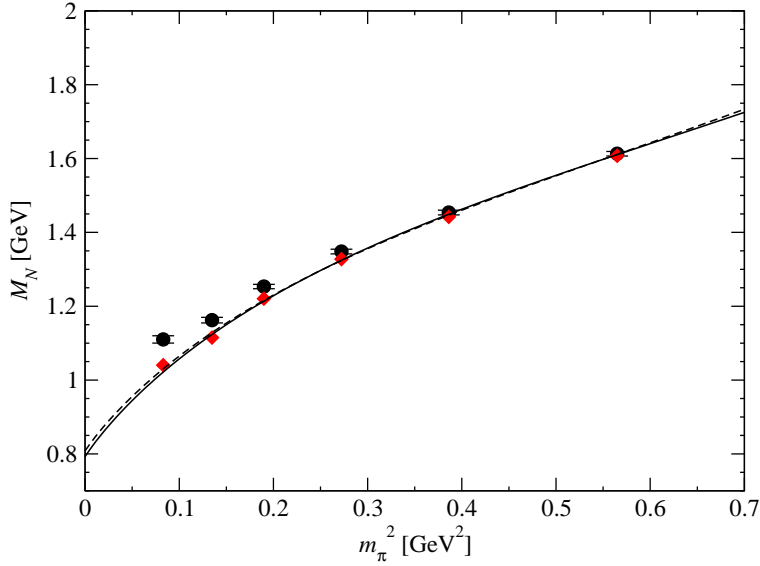}
  }
  \caption{Chiral fit of the corrected data (diamonds).
    Solid and dashed curves represent the fits using
    5 and 6 heaviest data points, respectively.
    For a reference, we also show the raw data (circles).}
  \label{fig:fit_FSE_FIT0}
\end{figure}

\begin{table}[tbp]
\centering
\begin{tabular}{ccccc}
\hline
   & $M_0$[GeV] &  $c_1$[GeV$^{-1}$] & $e_1^r(\mu)$ 
   & $\chi^2/{\mathrm{d.o.f.}}$ \\
   \hline
   Fit 0a(5pt)  &0.793(15) &$-$1.04(3) &2.68(15) &1.86 \\
   Fit 0a(6pt)  &0.808(13) &$-$1.02(3) &2.79(13) &1.82 \\
   \hline
 \end{tabular}
 \caption{
   Chiral fit parameters for the finite volume corrected lattice data.
   Results using all 6 data points ``Fit 0a (6pt)'' and those using
   5 heaviest data points ``Fit 0a (5pt)'' are listed.
   The fit function is (\ref{eq:p^3+higher}) with a fixed axial
   coupling $g_A$ = 1.267.
 }
 \label{tab:fit_FSE_hybrid}
\end{table}

\begin{figure}[tbp]
  \centering
  \rotatebox{0}{
    \includegraphics[width=10cm,clip]{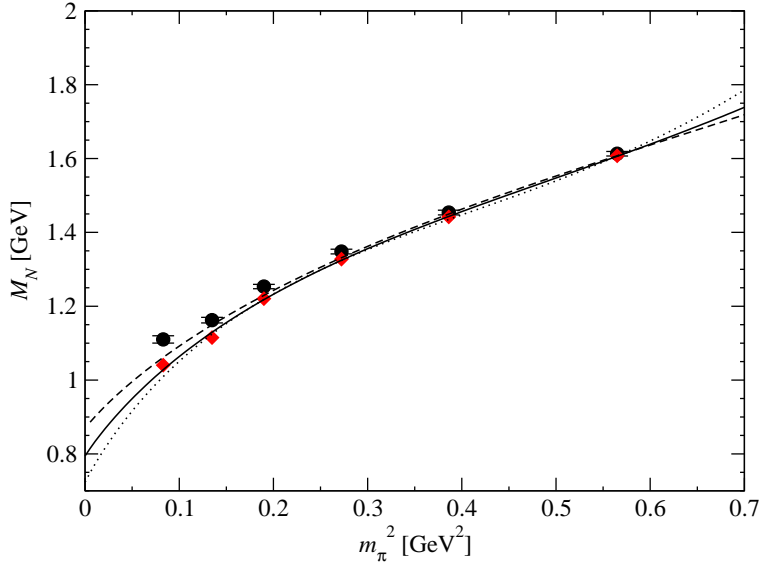}
  }
  \caption{
    Chiral fit with the $\mathcal{O}(p^4)$ formula plus the finite
    volume effect $\Delta_a+\Delta_b$.
    Dashed, solid and dotted curves represent the fit results in the infinite
    volume from the Fit Ia, II and III, respectively.
    For a reference, we show the finite volume corrected data points (diamonds).
  }
  \label{fig:fit_FSE_6pt}
\end{figure}

In the following we make the following two different 
analyses for the finite volume effect.
\begin{enumerate}
\item 
  We correct the data for the finite volume effect using the above
  formula. 
  For the input parameters $M_0$, $g_A$, $c_i$ ($i$ = 1--3), we use the
  nominal values ($M_0$ = 0.87~GeV, $g_A$ = 1.267, 
  $c_1=-1.0$ GeV$^{-1}$, $c_2=3.2$ GeV$^{-1}$, and $c_3=-3.4$ GeV$^{-1}$).
  The size of the finite volume corrections varies from $-0.3$\%
  (heaviest) to $-4.0$\% (second lightest) and $-6.5$\% (lightest).
  The chiral fit is then made for the corrected data points using the
  simplified $O(p^3)$ formula (\ref{eq:p^3+higher}) for heaviest 5 or
  6 heaviest data points. 
  The result is shown in Figure~\ref{fig:fit_FSE_FIT0} and the fit
  parameters are listed in Table~\ref{tab:fit_FSE_hybrid}.
  After correcting the finite volume effect, there is a 5-8\% 
  decrease in $M_0$ and 4-7\% increase in the magnitude of the slope 
  in the chiral limit $|c_1|$.
  The results of the fits with 5 or 6 data points are consistent
  with each other.
\item 
  We fit the data with the fit functions including the finite volume
  corrections, {\it i.e.}
  at $\mathcal{O}(p^3)$ the function is (\ref{eq:p^3}) plus $\Delta_a$
  (Fit II);
  at $\mathcal{O}(p^4)$ the function is (\ref{eq:p^4}) plus
  $\Delta_a+\Delta_b$ (Fit III).
  Figure~\ref{fig:fit_FSE_6pt} shows the fit curves after subtracting
  the finite volume piece $\Delta_a$ or $\Delta_a+\Delta_b$,
  which consistently run through the finite volume corrected data points. 
  The fit parameters are listed in Table~\ref{tab:FSE_p4}. 
  We find that after taking the finite volume effect into account
  $M_0$ decreases by 3-8\% and $|c_1|$ increases by 9\%.  
  The 5-point and 6-point fits are consistent with each other within
  two standard deviations.
\end{enumerate}
Comparing the fit parameters obtained with
(Tables~\ref{tab:fit_FSE_hybrid} and \ref{tab:FSE_p4}) and without
(Table~\ref{tab:ChPT}) the finite volume corrections, we observe that
the deviation due to the finite volume effect is smaller than the
uncertainty of the fit forms.

\begin{table}[tbp]
  \centering
  \begin{tabular}{cccccccc}
    \hline
    & $M_0$ & $c_1$ &$e_1^r(\mu)$ & $g_A$ & $c_2$ & $c_3$ 
    & $\chi^2/{\mathrm{d.o.f.}}$ \\ 
    & [GeV] & [GeV$^{-1}$] & [GeV$^{-3}$]  & &[GeV$^{-1}$] 
    & [GeV$^{-1}$]  &  \\ 
    \hline
    Fit Ia(5pt)   &0.852(15)&$-$0.90(3) &3.18(16)  &[1.267] 
    & -    & -        &1.81 \\
    Fit Ia(6pt)   &0.870(13)&$-$0.88(2) &3.31(13)  &[1.267] 
    & -    & -        &1.86 \\
    Fit II(5pt)   &0.778(12)&$-$1.08(2) &1.70(13)  &[1.267] 
    &[3.2] &[$-$3.4]  &1.19 \\
    Fit II(6pt)   &0.794(10)&$-$1.06(1) &1.83(11)  &[1.267] 
    &[3.2] &[$-$3.4]  &1.76 \\
    Fit III(5pt)  &0.694(12)&$-$1.35(2) &0.84(14)  &[1.267] 
    &[3.2] &[$-$4.7]  &0.24 \\
    Fit III(6pt)  &0.723(10)&$-$1.32(1) &1.10(12)  &[1.267] 
    &[3.2] &[$-$4.7]  &3.37 \\
    \hline
  \end{tabular}
  \caption{
    Results from the chiral fit including the finite volume
    corrections.
    The finite volume effects are included to $\Delta_a$ for the Fit
    Ia (at $\mathcal{O}(p^3)$) and to $\Delta_a+\Delta_b$ for the Fit
    II and III (at $\mathcal{O}(p^4)$).
  }
\label{tab:FSE_p4}
\end{table}

There is also a finite volume effect due to fixing the topological
charge in our simulation. 
This can be estimated using ChPT as in \cite{Brower:2003yx,Aoki:2007ka}.
The estimated corrections are fairly small, $-$(0.3--0.7)\% depending
on the quark mass.
Compared to the statistical error and the conventional finite volume
effect, we can safely neglect the fixed topology effect.
 
\subsection{Nucleon sigma term}

Using the fits in the previous subsections we obtain the nucleon sigma term 
by differentiating the nucleon mass with respect to the quark mass as
\begin{equation}
  \sigma_{\pi N} = \sum_{q=u,d}
  m_q\left.\frac{dM_N}{dm_q}\right|_{m_q=m_{ud}}.
\end{equation}
Since the value of the physical up and down quark mass is very small, 
we may extract the physical value using the leading-order ChPT relation 
\begin{equation}
  \sigma_{\pi N} = m_\pi^2 \left.\frac{dM_N}{dm_\pi^2}
  \right|_{m_\pi=135~\mathrm{MeV} }.
\end{equation}
Table~\ref{tab:sigma_FSE} shows the results from the several fit forms
with and without the finite volume corrections (FVCs).

\begin{table}[tbp]
  \centering
  \begin{tabular}{c|c|cc}
    \hline
    &  w/o FVCs &  \multicolumn{2}{c}{w/ FVCs} \\
    \hline
    & 5 pt  & 5 pt & 6 pt \\
    \hline
    Fit 0a  & 52.2(1.8) & 56.7(1.8)  & 55.1(1.5)  \\
    Fit Ia  & 45.1(1.7) & 48.9(1.7)  & 47.2(1.5)  \\
    Fit II  & 56.5(1.2) & 59.5(1.2)  & 58.2(1.0)  \\
    Fit III & 71.8(1.2) & 75.1(1.2)  & 72.7(1.0)  \\
   \hline
  \end{tabular}
  \caption{
    Nucleon sigma term $\sigma_{\pi N}$ [MeV]
    with and without the finite volume corrections (FVCs).
  } 
  \label{tab:sigma_FSE}
\end{table}

Due to the curvature observed in
Figures~\ref{fig:ChPT}--\ref{fig:fit_FSE_6pt}, that is largely
explained by the non-analytic term $m_\pi^3$ in the BChPT formulae, 
$\sigma_{\pi N}$ is enhanced toward the chiral limit.
Compared with the value at around the strange quark mass, 
$\sigma_{\pi N}$ is about three times larger, depending on the details
of the fit ansatz.

The largest uncertainty comes from the chiral extrapolation.
In fact, the Fit III gives significantly larger value of 
$\sigma_{\pi N}$ than those of other fit ansatz.
It is expected from the plot of chiral extrapolation,
Figure~\ref{fig:ChPT}, where the Fit III (dot-dashed curve) shows a
steeper slope near the chiral limit. 
The finite volume effect is a sub-leading effect, which is about 9\%.
We take the Fit 0a ($g_A$ fixed, FVCs not included) as our best fit,
and take the variation with fit ansatz and FVCs as an estimate of the
systematic error. 
We obtain
\begin{equation}
  \sigma_{\pi N} = 52(2)_{\rm stat}(^{+20}_{-\ 7})_{\rm
    extrap}(^{+5}_{-0})_{\rm FVE}
  \mathrm{~MeV}, 
  \label{eq:result-sigma}
\end{equation}
where the errors are the statistical and the systematic due to 
the chiral extrapolation (extrap) and finite volume effect (FVE).
This result is in good agreement with the phenomenological analysis
based on the experimental data at the Cheng-Dashen point
$\sigma_{\pi N}=55\sim 75$~MeV, which is discussed in the
Introduction.

\section{Analysis of the partially quenched data points}
\label{sec:PQChPT}

\subsection{Fits with partially quenched ChPT formula}
As described in Section~\ref{sec:Method}, the partial derivatives in
terms of the valence and sea quark masses, $m_{\mathrm{val}}$ and
$m_{\mathrm{sea}}$ respectively, are necessary in order to
extract the connected and disconnected-diagram contributions
separately, hence to obtain the strange quark content $y$ defined in
(\ref{eq:y}). 
It is possible with the lattice data in the so-called partially
quenched set-up, {\it i.e.} the valence quark mass is taken
differently from the sea quark mass.
Since the enhancement of $\sigma_{\pi N}$ towards the chiral limit is
essential for reliable determination of the nucleon sigma term, we
should use the chiral perturbation theory formula for baryons in
partially quenched QCD, which is available for two-flavor QCD
\cite{Chen:2001yi,Beane:2002vq}. 
At $O(p^3)$, it reads
\begin{eqnarray}
  \label{eq:PQChPT}
  M_N & = &  B_{00} + B_{10}(m_\pi^{vv})^2 +B_{01}(m_\pi^{ss})^2
  \nonumber \\
  & &  -\frac{1}{16\pi f_\pi^2}
       \Bigl\{ \frac{g_A^2}{12}
	\left[ -7(m_\pi^{vv})^3 +16 (m_\pi^{vs})^3 
               +9 m_\pi^{vv}(m_\pi^{ss})^2 )\right]  \nonumber \\
  & &  \quad\quad\quad\quad \
      + \frac{g_1^2}{12}
	\left[ -19(m_\pi^{vv})^3 +10 (m_\pi^{vs})^3 
              +9 m_\pi^{vv}(m_\pi^{ss})^2 )\right]  \nonumber \\
  & &  \quad\quad\quad\quad \
      + \frac{g_1 g_A}{3}
	\left[ -13(m_\pi^{vv})^3 + 4(m_\pi^{vs})^3 
               +9 m_\pi^{vv}(m_\pi^{ss})^2 )\right]
	\Bigr\} \nonumber\\
  & & +B_{20}(m_\pi^{vv})^4  
       +B_{11}(m_\pi^{vv})^2 (m_\pi^{ss})^2
       +B_{02}(m_\pi^{ss})^4 ,
\end{eqnarray}
where $m_\pi^{vv}$, $m_\pi^{vs}$, and $m_\pi^{ss}$ denote the pion
mass made of valence-valence, valence-sea, and sea-sea quark
combinations, respectively.
At this order of the chiral expansion, one can rewrite this formula in
terms of $m_{\mathrm{val}}$ and $m_{\mathrm{sea}}$ using the
leading-order relations
$(m_\pi^{vv})^2=2Bm_{\mathrm{val}}$,
$(m_\pi^{vs})^2=B(m_{\mathrm{val}}+m_{\mathrm{sea}})$, and
$(m_\pi^{ss})^2=2Bm_{\mathrm{sea}}$.
The parameter $B$ is determined as $B_0$ = 1.679(4)~GeV 
through the ChPT analysis of pion mass \cite{Noaki:2007es}.
The coupling constant $g_A$ represents the nucleon axial charge as
before, and $g_1$ is another axial-vector coupling characterizing the
coupling to the $\eta$ meson.
They are related to the standard $F$ and $D$ parameters of BChPT
as $g_A=F+D$ and $g_1=2(F-D)$.
Numerically, the values of $F$ and $D$ are obtained from the hyperon 
decay as $F=0.52(4)$ and $D=0.85(6)$ (see \cite{Luty:1993gi}, for
instance), which imply $g_1=-0.66(14)$.
In the following, whenever we need nominal values of $g_A$ and $g_1$,
we set $g_A=1.267$ and $g_1=-0.66$.

Strictly speaking, there are also contributions from the decuplet
baryons. 
In our analysis we have integrated out the Delta resonance 
and expand the contribution in terms of $(m_{\pi}/\Delta)^2$
with $\Delta=m_{\Delta}-M_N$. 
Then these contributions can be absorbed into the analytic terms in
(\ref{eq:PQChPT}). 

We fit the quark mass dependence of the nucleon mass with the
partially quenched ChPT formula (\ref{eq:PQChPT}).
The independent fit parameters are
$B_{00}$, $B_{01}$, $B_{10}$, $B_{11}$, $B_{20}$, $B_{02}$, $g_1$, and
$g_A$. 
Instead of making all these parameters free, we also attempt a fit
with fixed $g_A$ and $g_1$ (Fit PQ-a), a fit with fixed $g_A$ 
(Fit PQ-b). 
The fit with all the free parameters is called the Fit PQ-c.

\begin{table}[tbp]
  \begin{center}
    \begin{tabular}{cccccccccc}
      \hline
      & $B_{00}$ & $B_{01}$ & $B_{10}$ & $B_{11}$ & $B_{20}$ & $B_{02}$ 
      & $g_1$ & $g_A$ & $\chi^2/{\mathrm{d.o.f.}}$\\ 
      & [GeV] & [GeV$^{-1}$] & [GeV$^{-1}$] & [GeV$^{-3}$] &
      [GeV$^{-3}$] & [GeV$^{-3}$] &  &  & \\ 
      \hline
      Fit PQ-a  & 0.87(2) & 0.47(10)  & 3.37(3) & $-$0.94(2) & 3.77(2)
        & 0.17(15) & [$-$0.66] &  [1.267] & 1.82 \\ 
      Fit PQ-b  & 0.86(2) & 1.13(11)  & 2.71(4)  & 0.97(8) & 1.81(11)
        & 0.25(15) & $-$0.378(14) & [1.267] & 1.28 \\ 
      Fit PQ-c  & 0.92(4)& 0.76(23)  & 1.98(39) & 0.43(31) & 0.95(43) 
        & $-$0.03(23) & $-$0.29(5) & 0.93(22) & 1.28 \\ 
      \hline     
    \end{tabular}
    \caption{
      Fit results with the partially quenched ChPT formula.
      The values sandwiched as [$\cdots$] mean the input in the fit.
    }
    \label{tab:PQfitpara}
  \end{center}
\end{table}

\begin{figure}[tbp]
  \begin{center}
    \includegraphics[width=10cm,clip]{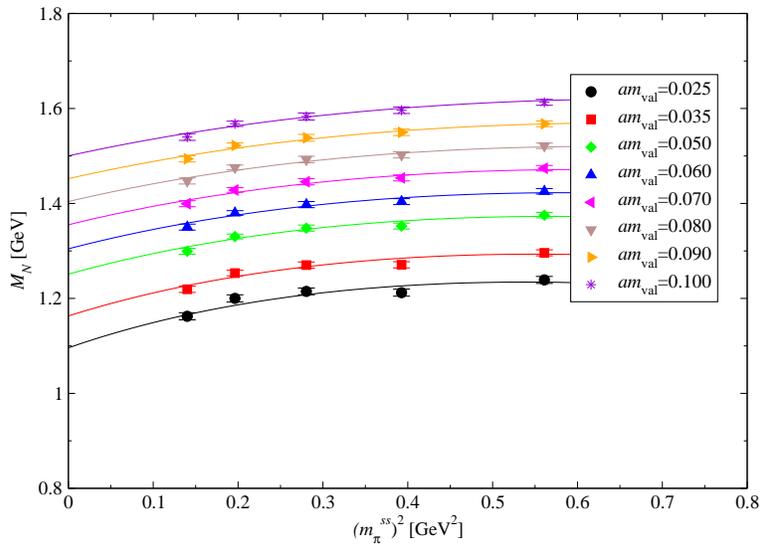}
  \end{center}
  \caption{Partially quenched nucleon masses and fit curves (Fit PQ-b).}
  \label{fig:PQChPT_ver2}
\end{figure}%

\begin{figure}[tbp]
  \begin{center}
    \rotatebox{0}{
      \includegraphics[width=10cm,clip]{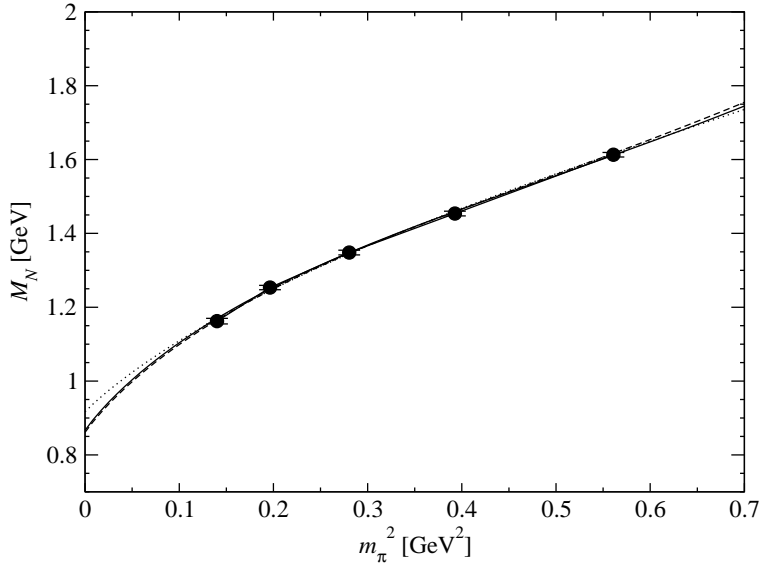}
    }
    \caption{Result of the PQChPT fit for unitary points.
     Solid, dashed and dotted curves represent the fit results 
     from the Fit PQ-a, PQ-b and PQ-c, respectively.}
    \label{fig:PQChPT} 
  \end{center}
\end{figure}%

Figure~\ref{fig:PQChPT_ver2} demonstrates the result of the partially
quenched ChPT fit.
It shows the sea quark mass dependence at eight fixed valence quark
masses.
Data are nicely fitted with the formula (\ref{eq:PQChPT}).
The fit results are listed in Table~\ref{tab:PQfitpara}.
All the parameters are well determined except for the term 
$B_{02}(m_\pi^{ss})^4$, for which the data do not have enough
sensitivity. 
Finite volume corrections are not taken into account.

By reducing the parameters to the case of the unitary point
$m_{\mathrm{val}}=m_{\mathrm{sea}}$
($m_0=B_{00}$, $c_1=-(B_{01}+B_{10})/4$, $e_1=B_{20}+B_{11}+B_{02}$),
it is easy to see that the results from Fits PQ-a and PQ-b are
consistent with the Fit 0a for the unitary points.
Figure~\ref{fig:PQChPT} shows the reduction to the unitary point.
The value of the nucleon sigma term $\sigma_{\pi N}$ obtained from
this reduced fit parameters is 53.3(1.8), 53.2(1.9) and 41.3(6.6)~MeV
for the Fits PQ-a, PQ-b, and PQ-c, respectively.
These values are in good agreement with our analysis of the unitary
points (\ref{eq:result-sigma}).

Another important observation from Table~\ref{tab:PQfitpara} is that
the Fit PQ-c, for which $g_A$ is a free parameter, gives much better
constrained $g_A$ than the Fit 0b of the unitary points. 
This is because the partially quenched analysis uses much more data
points: 40 data points compared to 5 in the unitary case.
It is remarkable that both $g_A$ and $g_1$ can be determined with
reasonable accuracy.

\subsection{Sea quark content of the nucleon}
Once the valence and sea quark mass dependence is identified using
the formula (\ref{eq:PQChPT}), we can obtain the partial derivatives
with respect to $m_{\mathrm{val}}$ and $m_{\mathrm{sea}}$ to obtain the
connected and the disconnected contribution to the nucleon sigma
term $\sigma_{\pi N}$ as defined in (\ref{eq:deriv_mN_val}) and
(\ref{eq:deriv_mN_sea}). 

\begin{figure}[tbp]
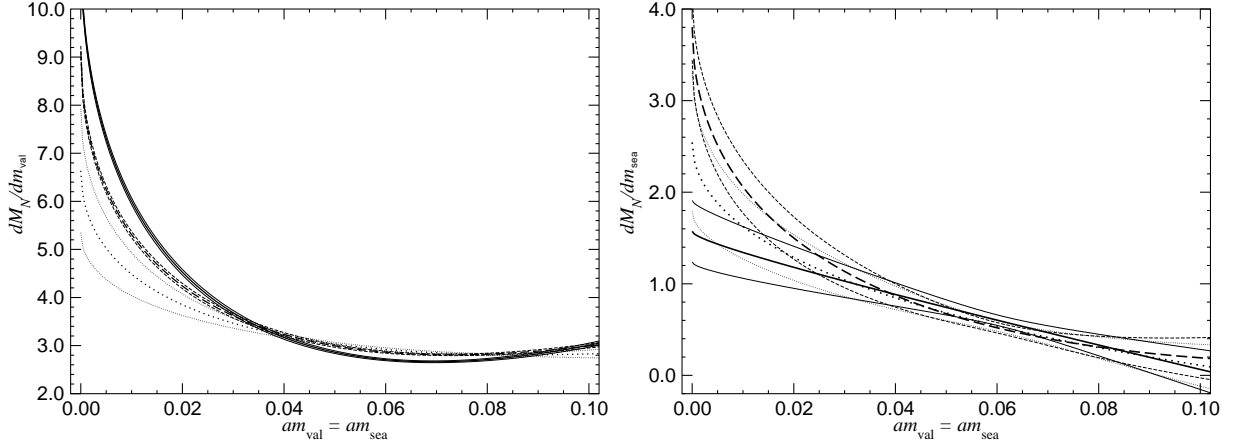

  \rotatebox{0}{
    \includegraphics[width=8cm,clip]{./figure/dmdv.eps}
    \includegraphics[width=8cm,clip]{./figure/dmds.eps}
  }
  \caption{
    Connected (left) and disconnected (right) contributions 
    to the nucleon sigma term evaluated at $m_{\mathrm{val}}=m_{\mathrm{sea}}$.
    Solid, dashed and dotted curves   
    represent the results from the Fit PQ-a, PQ-b and PQ-c, 
    respectively (thick lines).   
    The error curves are represented by the thin lines. 
  }
  \label{fig:conn_and_disc}
\end{figure}%

\begin{table}[tbp]
  \begin{center}
    \begin{tabular}{c|ccc|ccc}
      \hline
      & $a m_q$
      & ${\displaystyle \frac{\partial M_N}{\partial m_{\rm val}}}$ 
      & ${\displaystyle \frac{\partial M_N}{\partial m_{\rm sea}}}$ 
      & $a m_q$ 
      & ${\displaystyle \frac{\partial M_N}{\partial m_{\rm val}}}$ 
      & ${\displaystyle \frac{\partial M_N}{\partial m_{\rm sea}}}$ 
      \\
      \hline
      Fit PQ-a &0.0034 & 7.92(8)  & 1.47(32) & 0.084 & 2.75(3)  & 0.28(14)\\
      Fit PQ-b &0.0034 & 6.68(8)  & 2.72(33) & 0.084 & 2.84(3)  & 0.28(14)\\
      Fit PQ-c &0.0034 & 5.27(75)  & 1.99(50) & 0.084 & 2.81(3)  & 0.26(14)\\
      \hline
    \end{tabular}
    \caption{
      Connected and disconnected contributions to the nucleon sigma term,
      evaluated at the average up and down quark mass $am_q=0.0034$
      and at the physical strange quark mass $am_q=0.084$.
    }
    \label{tab:conn_disc_sigma}
  \end{center}
\end{table}

Figure~\ref{fig:conn_and_disc} shows the partial derivatives with
respect to $m_{\mathrm{val}}$ (left panel) and to $m_{\mathrm{sea}}$
(right panel) evaluated at the unitary points $m_{\mathrm{val}}=m_{\mathrm{sea}}$.
In the plots, the fit results are plotted as a function of 
$m_{\mathrm{val}}=m_{\mathrm{sea}}$.
For both contributions, we clearly find an enhancement towards the
chiral limit.
Results with different fit ansatz show slight disagreement near the
chiral limit, which indicate the size of the systematic uncertainty.

Numerical results at the average up and down quark mass and at the
physical strange quark mass are listed in Table~\ref{tab:conn_disc_sigma}.
The values in the lattice unit $am_{ud}=0.0034(1)$ and $am_s=0.084(2)$ are
determined from a partially quenched analysis of the meson spectrum
\cite{Noaki:2007es}. 

\begin{figure}[tbp]
  \begin{center}
    \rotatebox{0}{
      \includegraphics[width=10cm,clip]{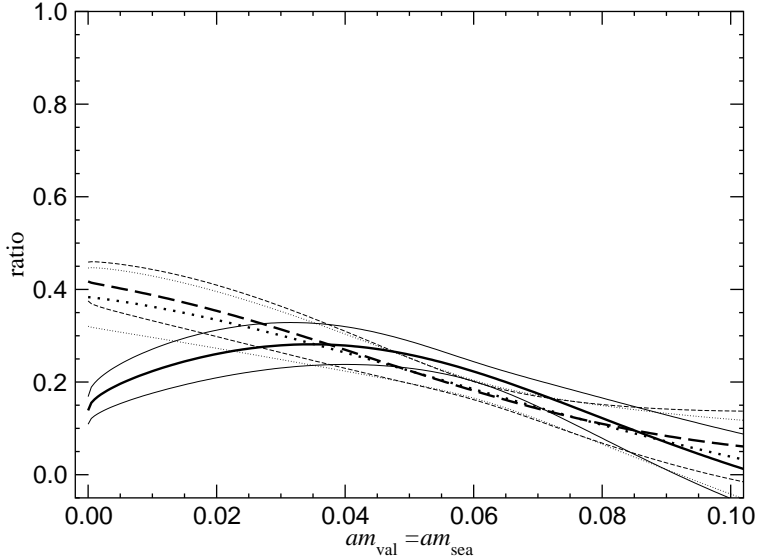}
    }
  \end{center}
  \caption{
    Ratio of the 
    disconnected and connected contribution to the sigma term
    for unitary points ($m_{\rm sea}=m_{\rm val}$).
    Solid, dashed and dotted curves   
    represent the results from the Fit PQ-a, PQ-b and PQ-c, 
    respectively (thick lines).   
    The error curves are represented by the thin lines. 
  } \label{fig:ratio}
\end{figure}%

Figure~\ref{fig:ratio} shows the ratio of the disconnected and
connected contribution to the sigma term 
$\langle N|(\bar{u}u+\bar{d}d)|N\rangle_{\mathrm{disc}}/
\langle N|(\bar{u}u+\bar{d}d)|N\rangle_{\mathrm{conn}}$
evaluated at the unitary points $m_{\rm sea}=m_{\rm val}$.
We find that the sea quark content of the nucleon is less than 0.4 for
the entire quark mass region in our study, so that the
valence quark content is the dominant contribution to the sigma term.
This is in striking contrast to the previous lattice results 
in which the sea quark content equal to or even larger than the
valence quark content was found.

\subsection{A semi-quenched estimate of the strange quark 
content}
Rigorously speaking, it is not possible to extract the strange quark
content 
$\langle N | \bar{s}s | N \rangle$ within two-flavor QCD.
The problem is not just the strange quark loop is missing, 
but it is not possible to evaluate the disconnected contribution 
at the strange quark mass 
while sending the sea and valence quark masses to the
physical up and down quark mass with the partial derivatives 
within (partially quenched) two-flavor QCD.
For the final result, therefore, we should wait for a 2+1-flavor QCD
simulation, which is in progress \cite{Hashimoto:2007vv}. 
Instead, in this work, we provide a ``semi-quenched'' estimate of the
strange quark content assuming that the disconnected contribution  
gives a good estimate of the strange quark effect when evaluated 
at the strange quark mass for both $m_{\mathrm{val}}$ and
$m_{\mathrm{sea}}$. 

We define our semi-quenched estimate of the parameter y 
as the ratio of the strange quark content 
(disconnected contribution at
$m_{\mathrm{val}}=m_{\mathrm{sea}}=m_s$) to the up and down quark
contributions (connected plus disconnected contributions at
$m_{\mathrm{val}}=m_{\mathrm{sea}}=m_{ud}$) following 
Ref.~\cite{Michael:2001bv}.
Taking the result from the Fit PQ-b as a best estimate,  
we obtain the 
parameter $y$ as 
\begin{equation}
  \label{eq:y_result}
  y^{N_f=2} = 0.030(16)_{\rm stat}(^{+6}_{-8})_{\rm extrap}(^{+1}_{-2})_{m_s},
\end{equation}
where the errors are statistical, the systematic errors from chiral
extrapolation and from the uncertainty of $m_s$.  
The chiral extrapolation error  for the strange quark content 
is estimated by the differences of the results of Fit PQ-a,b and c, 
while that for the up and down quark content is estimated by the 
differences of the results of Fit 0, I, II and III. We also note that 
there may be an additional $\sim$10\% error from finite volume effect 
as discussed in Section~\ref{sec:BChPT}, but it is much smaller than 
the statistical error in our calculation.

\section{Discussion}
\label{sec:Discussion}

We found that the disconnected contribution to the sigma term is much 
smaller than the previous lattice calculations with the Wilson-type
fermions $y\simeq 0.36\sim 0.66$ 
\cite{Fukugita:1995ba,Dong:1996ec,Gusken:1998wy}
(except for \cite{Michael:2001bv} as explained below).
The authors of \cite{Michael:2001bv} found that the naive calculation
with the Wilson-type fermions may over-estimate the sea quark contents 
due to the additive mass shift and the sea quark mass dependence of
the lattice spacing. 
The key observation is that the additive mass shift is large depending
significantly on the sea quark mass.
Therefore, in order to obtain the derivative (\ref{eq:deriv_sea}) one
must subtract the unphysical contribution from the additive mass
shift.
This problem remains implicitly in the quenched calculations, since
the derivative must be evaluated at the value of the valence quark
mass even when the sea quark mass is sent to infinity.
(There is of course the more fundamental problem in the quenched
calculations due to the missing sea quark effects.)

Another problem is in the conventional scheme of setting the lattice
scale in unquenched simulations.
In many dynamical fermion simulations, the lattice spacing is set
(typically using the Sommer scale $r_0$) at each sea quark mass, or in
some cases, the bare lattice coupling $\beta$ is tuned to yield a given
value of $r_0$ independent of the sea quark mass.
This procedure defines a renormalization scheme that is mass
dependent, because the quantity $r_0$ could have physical sea quark
mass dependence.
Since the partial derivative (\ref{eq:deriv_sea}) is defined in a mass
independent scheme, {\it i.e.} the coupling constant does not depend
on the sea quark mass, one has to correct for the artificial sea quark
mass dependence through $r_0$ when one calculates the nucleon sigma
term. 
Combining these two effects, the authors of \cite{Michael:2001bv}
found that their unsubtracted result $y=0.53(12)$ is substantially
reduced and becomes consistent with zero: $y=-0.28(33)$.
The conclusion of this analysis is that the previous lattice
calculations giving the large values of $y$ suffered from the large
systematic effect, hence should not be taken at their face values.

Our calculation using the overlap fermion is free from these
artifacts. 
The additive mass shift is absent because of the exact chiral
symmetry of the overlap fermion.
The lattice spacing is kept fixed in our analysis at a fixed bare
lattice coupling constant.
We confirmed that this choice gives a constant value of the
renormalized coupling constant in the (mass independent)
$\overline{\mathrm{MS}}$ scheme through an analysis of current-current
correlators \cite{Shintani:2008qe}.
Therefore, the small value of $y$ obtained in our analysis
(\ref{eq:y_result}) provides a much more reliable estimate than the previous
lattice calculations.

\section{Summary}
\label{sec:Summary}
We study the nucleon sigma term in two-flavor QCD simulation on the
lattice with exact chiral symmetry.
Fitting the quark mass dependence of the nucleon mass using the
formulae from Baryon Chiral Perturbation Theory (BChPT), we obtain
$\sigma_{\pi N} = 53(2)(^{+21}_{-\ 7})$~MeV, where our estimates of
systematic errors are added in quadrature.
This is consistent with the canonical value in the previous
phenomenological analysis.
Owing to the exact chiral symmetry, our lattice calculation is free
from the large lattice artifacts coming from the additive mass shift
present in the Wilson-type fermion formulations.

We also estimate the strange quark content of nucleon.
From an analysis of partially quenched lattice data, we find that the
sea quark content of the nucleon is less than 0.4 for the entire quark
mass region in our study.
The valence quark content is in fact the dominant contribution to 
the sigma term.
Taking account of the enhancement of 
$\langle N|(\bar{u}u+\bar{d}d)|N\rangle$ near the chiral limit, 
the parameter $y$ is most likely less than 0.05 in contrast to the 
previous lattice calculations.

By directly calculating the disconnected diagram we may obtain further
information. 
For instance, the effect of the strange quark loop on the dynamical
configurations with light up and down quarks can be extracted.
Such a calculation is in progress using the all-to-all quark propagators on
the lattice.
Another obvious extension of this work is the calculation including
the strange quark loop in the vacuum.
Simulations with two light and one strange dynamical overlap quarks
are on-going \cite{Hashimoto:2007vv}.

\begin{acknowledgments}
We would like to thank R.~Kitano for a suggestion to work on this
subject. 
We acknowledge K. Aoki, M. Nojiri, J. Hisano for fruitful discussions. 
We also thank W. A. Bardeen and M. Peskin for discussions and 
crucial comments. Special thanks to D. Jido and T. Kunihiro for useful 
discussions and informing us about the recent developments in nucleon 
sigma term in chiral perturbation theory. We also thank S. Aoki for
careful reading of the manuscript and crucial comments.
 
We acknowledge the international 'molecule' visitor program supported
by the Yukawa International Program for Quark-Hadron Sciences (YIPQS),
where intensive discussions with the visitors helped to proceed this
work. 

The main numerical calculations were performed on IBM System Blue Gene
Solution at High Energy Accelerator Organization (KEK) under support
of its Large Scale Simulation Program (No.~07-16).
We also used NEC SX-8 at Yukawa Institute for Theoretical Physics
(YITP), Kyoto University and at Research Center for Nuclear Physics
(RCNP), Osaka University. 
The simulation also owes to a gigabit network SINET3 supported by
National Institute of Informatics for efficient data transfer through
Japan Lattice Data Grid (JLDG).
This work is supported in part by the Grant-in-Aid of the
Ministry of Education (Nos. 18034011, 18340075, 18740167, 
19540286, 19740121, 19740160, 20025010, 20039005).
The work of HF is supported by Nishina Memorial Foundation.
\end{acknowledgments}


\begin{thebibliography}{99}

\bibitem{Griest:1988yr}
  K.~Griest,
  Phys.\ Rev.\ Lett.\  {\bf 61}, 666 (1988).

\bibitem{Griest:1988ma}
  K.~Griest,
  Phys.\ Rev.\  D {\bf 38}, 2357 (1988)
  [Erratum-ibid.\  D {\bf 39}, 3802 (1989)].

\bibitem{Bottino:1999ei}
  A.~Bottino, F.~Donato, N.~Fornengo and S.~Scopel,
  Astropart.\ Phys.\  {\bf 13}, 215 (2000)
  [arXiv:hep-ph/9909228].

\bibitem{Ellis:2003cw}
  J.~R.~Ellis, K.~A.~Olive, Y.~Santoso and V.~C.~Spanos,
  Phys.\ Lett.\  B {\bf 565}, 176 (2003)
  [arXiv:hep-ph/0303043].

\bibitem{Ellis:2005mb}
  J.~R.~Ellis, K.~A.~Olive, Y.~Santoso and V.~C.~Spanos,
  Phys.\ Rev.\  D {\bf 71}, 095007 (2005)
  [arXiv:hep-ph/0502001].

\bibitem{Baltz:2006fm}
  E.~A.~Baltz, M.~Battaglia, M.~E.~Peskin and T.~Wizansky,
  Phys.\ Rev.\  D {\bf 74}, 103521 (2006)
  [arXiv:hep-ph/0602187].

\bibitem{Ellis:2008hf}
  J.~Ellis, K.~A.~Olive and C.~Savage,
  Supersymmetric Dark
  arXiv:0801.3656 [hep-ph].

\bibitem{Cheng:1970mx}
  T.~P.~Cheng and R.~F.~Dashen,
  Phys.\ Rev.\ Lett.\  {\bf 26}, 594 (1971).

\bibitem{Pavan:2001wz}
  M.~M.~Pavan, I.~I.~Strakovsky, R.~L.~Workman and R.~A.~Arndt,
  PiN Newslett.\  {\bf 16}, 110 (2002)
  [arXiv:hep-ph/0111066].

\bibitem{Gasser:1990ce}
  J.~Gasser, H.~Leutwyler and M.~E.~Sainio,
  Phys.\ Lett.\  B {\bf 253}, 252 (1991).

\bibitem{Borasoy:1996bx}
  B.~Borasoy and U.~G.~Meissner,
  Annals Phys.\  {\bf 254}, 192 (1997)
  [arXiv:hep-ph/9607432].

\bibitem{Bernard:2007zu}
  V.~Bernard,
  Prog.\ Part.\ Nucl.\ Phys.\  {\bf 60}, 82 (2008)
  [arXiv:0706.0312 [hep-ph]].

\bibitem{Fukugita:1995ba}
M. Fukugita et~al.,
\newblock Phys. Rev. D51 (1995) 5319, hep-lat/9408002,
\newblock 

\bibitem{Dong:1996ec}
S.J. Dong, J.F. Lagae and K.F. Liu,
\newblock Phys. Rev. D54 (1996) 5496, hep-ph/9602259,
\newblock 

\bibitem{Gusken:1998wy}
SESAM, S. Gusken et~al.,
\newblock Phys. Rev. D59 (1999) 054504, hep-lat/9809066,
\newblock 

\bibitem{Michael:2001bv}
  C.~Michael, C.~McNeile and D.~Hepburn  [UKQCD Collaboration],
  Nucl.\ Phys.\ Proc.\ Suppl.\  {\bf 106}, 293 (2002)
  [arXiv:hep-lat/0109028].

\bibitem{Ali Khan:2001tx}
  A.~Ali Khan {\it et al.}  [CP-PACS Collaboration],
  Phys.\ Rev.\  D {\bf 65}, 054505 (2002)
  [Erratum-ibid.\  D {\bf 67}, 059901 (2003)]
  [arXiv:hep-lat/0105015].

\bibitem{Aoki:2002uc}
  S.~Aoki {\it et al.}  [JLQCD Collaboration],
  Phys.\ Rev.\  D {\bf 68}, 054502 (2003)
  [arXiv:hep-lat/0212039].

\bibitem{Ali Khan:2003cu}
  A.~Ali Khan {\it et al.}  [QCDSF-UKQCD Collaboration],
  Nucl.\ Phys.\  B {\bf 689}, 175 (2004)
  [arXiv:hep-lat/0312030].

\bibitem{Procura:2003ig}
  M.~Procura, T.~R.~Hemmert and W.~Weise,
  Phys.\ Rev.\  D {\bf 69}, 034505 (2004)
  [arXiv:hep-lat/0309020].

\bibitem{Procura:2006bj}
  M.~Procura, B.~U.~Musch, T.~Wollenweber, T.~R.~Hemmert and W.~Weise,
  Phys.\ Rev.\  D {\bf 73}, 114510 (2006)
  [arXiv:hep-lat/0603001].

\bibitem{Alexandrou:2008tn}
  C.~Alexandrou {\it et al.}  [European Twisted Mass Collaboration],
  Phys.\ Rev.\  D {\bf 78}, 014509 (2008)
  [arXiv:0803.3190 [hep-lat]].

\bibitem{Aoki:2008tq}
  S.~Aoki {\it et al.}  [JLQCD Collaboration],
  arXiv:0803.3197 [hep-lat].

\bibitem{Matsufuru:2007uc}
  H.~Matsufuru  [JLQCD Collaboration],
  PoS {\bf LAT2007}, 018 (2007)
  [arXiv:0710.4225 [hep-lat]].

\bibitem{Neuberger:1997fp}
  H.~Neuberger,
  Phys.\ Lett.\ B {\bf 417}, 141 (1998)
  [arXiv:hep-lat/9707022].

\bibitem{Neuberger:1998wv}
  H.~Neuberger,
  Phys.\ Lett.\ B {\bf 427}, 353 (1998)
  [arXiv:hep-lat/9801031].

\bibitem{WalkerLoud:2008bp}
  A.~Walker-Loud {\it et al.},
  arXiv:0806.4549 [hep-lat].

\bibitem{Fukaya:2006vs}
  H.~Fukaya, S.~Hashimoto, K.~I.~Ishikawa, T.~Kaneko, H.~Matsufuru,
  T.~Onogi and N.~Yamada
                  [JLQCD Collaboration],
  Phys.\ Rev.\  D {\bf 74}, 094505 (2006)
  [arXiv:hep-lat/0607020].

\bibitem{Noaki:2008iy}
  J.~Noaki {\it et al.} [JLQCD and TWQCD collaborations],
  arXiv:0806.0894 [hep-lat].

\bibitem{DeGrand:2004qw}
  T.~A.~DeGrand and S.~Schaefer,
  Comput.\ Phys.\ Commun.\  {\bf 159}, 185 (2004)
  [arXiv:hep-lat/0401011].

\bibitem{Jenkins:1990jv}
  E.~E.~Jenkins and A.~V.~Manohar,
  Phys.\ Lett.\  B {\bf 255}, 558 (1991).

\bibitem{Yao:2006px}
  W.~M.~Yao {\it et al.}  [Particle Data Group],
  J.\ Phys.\ G {\bf 33}, 1 (2006).

\bibitem{Brower:2003yx}
  R.~Brower, S.~Chandrasekharan, J.~W.~Negele and U.~J.~Wiese,
  Phys.\ Lett.\  B {\bf 560}, 64 (2003)
  [arXiv:hep-lat/0302005].

\bibitem{Aoki:2007ka}
  S.~Aoki, H.~Fukaya, S.~Hashimoto and T.~Onogi,
  Phys.\ Rev.\  D {\bf 76}, 054508 (2007)
  [arXiv:0707.0396 [hep-lat]].

\bibitem{Noaki:2007es}
  J.~Noaki {\it et al.}  [JLQCD Collaboration],
  PoS {\bf LAT2007}, 126 (2007)
  [arXiv:0710.0929 [hep-lat]].

\bibitem{Chen:2001yi}
  J.~W.~Chen and M.~J.~Savage,
  Phys.\ Rev.\  D {\bf 65}, 094001 (2002)
  [arXiv:hep-lat/0111050].

\bibitem{Beane:2002vq}
  S.~R.~Beane and M.~J.~Savage,
  Nucl.\ Phys.\  A {\bf 709}, 319 (2002)
  [arXiv:hep-lat/0203003].

\bibitem{Luty:1993gi}
  M.~A.~Luty and M.~J.~White,
  Phys.\ Lett.\  B {\bf 319}, 261 (1993).

\bibitem{Shintani:2008qe}
  E.~Shintani {\it et al.}  [for JLQCD Collaboration],
  arXiv:0806.4222 [hep-lat].

\bibitem{Hashimoto:2007vv}
  S.~Hashimoto {\it et al.}  [JLQCD collaboration],
  PoS {\bf LAT2007}, 101 (2007)
  [arXiv:0710.2730 [hep-lat]].


\end{thebibliography}
\end{document}